\documentclass[preprints,article,accept,pdftex,moreauthors]{Definitions/mdpi} 
\firstpage{1} 
\pubvolume{1}
\issuenum{1}
\articlenumber{0}
\pubyear{2022}
\copyrightyear{2022}
\datereceived{} 
\dateaccepted{} 
\datepublished{} 
\hreflink{https://doi.org/} 

\Title{From inclusive to semi-inclusive one-nucleon knockout in neutrino event generators}

\TitleCitation{From inclusive to semi-inclusive one-nucleon knockout in neutrino event generators}

\Author{Alexis Nikolakopoulos $^{1,}$*, Steven Gardiner $^{1}$, Afroditi Papadopoulou $^{2}$, Stephen Dolan $^{3}$ and Ra\'ul Gonz\'alez-Jim\'enez $^{4}$}

\AuthorNames{Alexis Nikolakopoulos, Steven Gardiner, Afroditi Papadopoulou, Stephen Dolan, Ra\'ul Gonz\'alez-Jim\'enez}

\AuthorCitation{Nikolakopoulos, A.; Gardiner, S.; Papadopoulou, A. ; Dolan S. ; Gonz\'alez-Jim\'enez, R.}

\address{%
$^{1}$ \quad Fermi National Accelerator Laboratory (Fermilab), P.O. Box 500, Batavia, IL 60510, USA. \\
$^{2}$ \quad Argonne National Laboratory (ANL), Lemont, IL, 60439, USA \\
$^{3}$ \quad European Organization for Nuclear Research (CERN), 1211 Geneva 23, Switzerland. \\
$^{4}$ \quad Grupo de F\'isica Nuclear, Departamento de Estructura de la Materia, F\'isica T\'ermica y Electr\'onica, Facultad de Ciencias F\'isicas, Universidad Complutense de Madrid and IPARCOS, CEI Moncloa, Madrid 28040, Spain. }

\corres{Correspondence: anikolak@fnal.gov; }

\abstract{
In neutrino event generators, for models for neutrino and electron scattering only inclusive cross sections are implemented.
When these models are used to describe a semi-inclusive cross section, the event generator attaches the hadron variables based on some assumptions.
In this work we compare the nucleon kinematics given by the method used in the GENIE event generator, e.g. in the implementation of the SuSAv2 model, to a fully unfactorized calculation using the relativistic distorted wave impulse approximation (RDWIA).
We focus on kinematics relevant to the $e4\nu$ analysis and show that observables obtained with RDWIA differ significantly from those of the approximate method used in GENIE, the latter should be considered unrealistic.
}

\begin{document}

\section{Introduction}
In recent years, accelerator-based neutrino experiments have performed measurements of the hadronic final-state in charged-current interactions. 
The hadron information is useful to distinguish between different interaction mechanisms and probe part of the nuclear momentum distribution through, for example, transverse kinematic imbalance~\cite{Lu:2015, Dolan:2018zye}. Additionally, a precise determination of outgoing nucleon kinematics, with suitable kinematic cuts leads to a more precise reconstruction of neutrino energy on an event-by-event basis~\cite{Gonzalez-Jimenez:2021ohu}.

The main challenge in accelerator-based neutrino experiments is that the incoming energy distribution is broad. This means that, to describe a $A(\nu_\mu,\mu p)X$ signal one has to account for a wealth of interaction mechanisms. At sufficiently high energy and momentum transfers, we could assume that the energy-momentum is absorbed by a single nucleon which reinteracts with the nucleus. These strong interactions are dubbed final-state interactions (FSI). Within this picture, inelastic FSI, where the energy gets distributed over many different final-states, is particularly important. The large phase-space covered in experiments, and the fact that the total energy of the residual system $X$ is not strongly constrained, make a quantum-mechanical description of all the possible coupled final-state channels intractable.
For this reason experiments deal with this problem using intranuclear cascade models (INC), which provide an explicit, albeit (semi-)classical, description of rescattering~\cite{NuWro:FSI, Hayato:NEUT,NEUT:2021EPJST, Salcedo88, GENIE, Isaacson:2022cwh, Dytman:2021ohr}.

For the following we consider the contribution of quasielastic interactions, where the exchanged boson is absorbed on a single nucleon that is excited to the continuum, to a 1-lepton 1-proton final state, e.g. to $A(\nu_\mu,\mu p)X$ or $A(e,e^\prime p)X$. 
Within the GENIE event generator the process is described by introducing this nucleon into the INC, which redistributes the strength over final-states in a unitary way. This means that after integration over all final-states the original inclusive cross section is recovered.
A number of models which describe the inclusive cross sections in the vicinity of the quasielastic peak, have been implemented in GENIE~\cite{Dolan:2021rdd, Dolan19, NIEVES2011, NIEVES2013, MM2013}. 
The caveat is that only inclusive cross sections are provided, the full kinematic and dynamic structure of the semi-inclusive cross section~\cite{Moreno:2014} is not available.
Therefore GENIE provides a procedure to attach the outgoing nucleon kinematics given some nuclear momentum distribution, in Ref.~\cite{Dolan19} discrepancies between this method and a microscopic calculation were discussed for flux-averaged neutrino cross sections.

In the following we present the algorithm used in GENIE to compute nucleon variables. We compare this result to fully unfactorized calculations with the relativistic distorted-wave impulse approximation (RDWIA). We focus on the $(e,e^\prime p)$ process for kinematics relevant to the $e4\nu$ analysis~\cite{CLAS:2021neh}.
 We show that the approximate procedure used in GENIE leads to significantly different observables than the RDWIA result.

\section{Results}
We use an unfactorized RDWIA calculation where the final-state nucleon is described in the real Energy-Dependent Relativistic Mean Field (ED-RMF) potential~\cite{Gonzalez-Jimenez19}.
The ED-RMF model yields an inclusive $(e,e^\prime)$ cross section that is practically the same as the SuSAv2 model for sufficient momentum transfer~\cite{Gonzalez-Jimenez:2019ejf}.
The main power of this approach is that it provides an exclusive cross section, unlike the SuSAv2 approach. 
In Ref.~\cite{Nikolakopoulos:2022qkq} the RDWIA with a real potential was used as input to the INC in the NEUT generator~\cite{NEUT:2021EPJST}. The results compare well with the T2K data of Ref.~\cite{T2K:2018rnz}. It was moreover found that, at sufficiently high nucleon energy ($T_N \gtrsim 100 \mathrm{MeV}$), the resulting exclusive cross sections agree with optical potential calculations.

\begin{figure}
\centering
\includegraphics[width=0.95\textwidth]{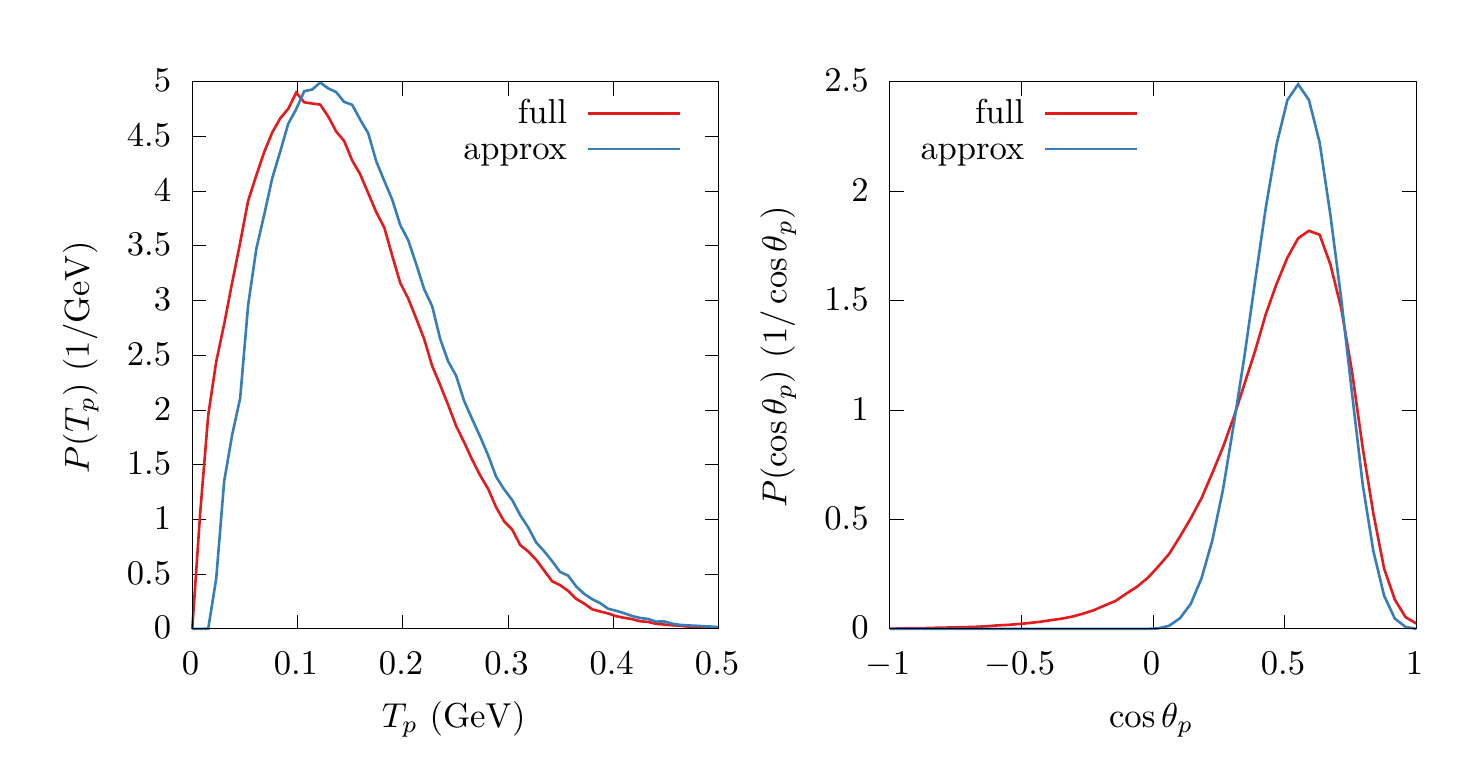}
\caption{Distributions of nucleon kinetic energy (left) and scattering angle with respect to the electron beam (right).  Results for scattering of $1.159~\mathrm{GeV}$ electrons off carbon. The red lines show the unfactorized RDWIA calculation using the ED-RMF potential, the blue lines show the result of the algorithm used in GENIE that uses the same inclusive cross section as input.}
\label{fig:TP_CTp_EDRMF}
\end{figure}

We have generated events for the process $e + A \rightarrow e^\prime + p + B$ with $E_e = 1.159~\mathrm{GeV}$. The events are distributed according to the $(Q^2)^2$-weighted cross section obtained in the RDWIA (for details see Refs.~\cite{Nikolakopoulos:2022qkq, Gonzalez-Jimenez:2021ohu}) 
\begin{equation}
P(E_l, \theta_l, T_N, \Omega_N) = \left(\frac{Q^2}{1~\mathrm{GeV}^2}\right)^2 \sum_{M_B} \frac{d^4 \sigma(E_e, M_B)}{\,d E_{e^\prime} \,d \cos\theta_{e^\prime} \,d \Omega_N}.
\end{equation}
The $(Q^2)^2$ weighting, which was introduced in the $e4\nu$ analysis of Ref.~\cite{CLAS:2021neh}, makes the kinematic dependence of the weighted cross section similar to the neutrino scattering case. The sum is over the invariant masses of the residual system, given by the RMF model as in Ref.~\cite{Nikolakopoulos:2022qkq}; we only include scattering with protons.
We impose cuts for the outgoing electron $40^{\circ} > \theta_{e^\prime} > 17^{\circ}$ and $E_{e^\prime} > 400~\mathrm{MeV}$, but include the full nucleon phase space. 

To test the approximate treatment, we replace the nucleon variables for every event by the ones produced by the algorithm used in GENIE, which is described in the following.
As the momentum distribution we use the local Fermi gas (FG) obtained from the nuclear density, also taken from GENIE,
\begin{equation}
\rho(r) = N\left(1+\left(\frac{r}{a}\right)^2\alpha \right)e^{\left(\frac{r}{a}\right)^2},
\end{equation}
with $a=1.69~\mathrm{fm}$ and $\alpha = 1.08$ for carbon.

\begin{figure}
\centering
\includegraphics[width=0.95\textwidth]{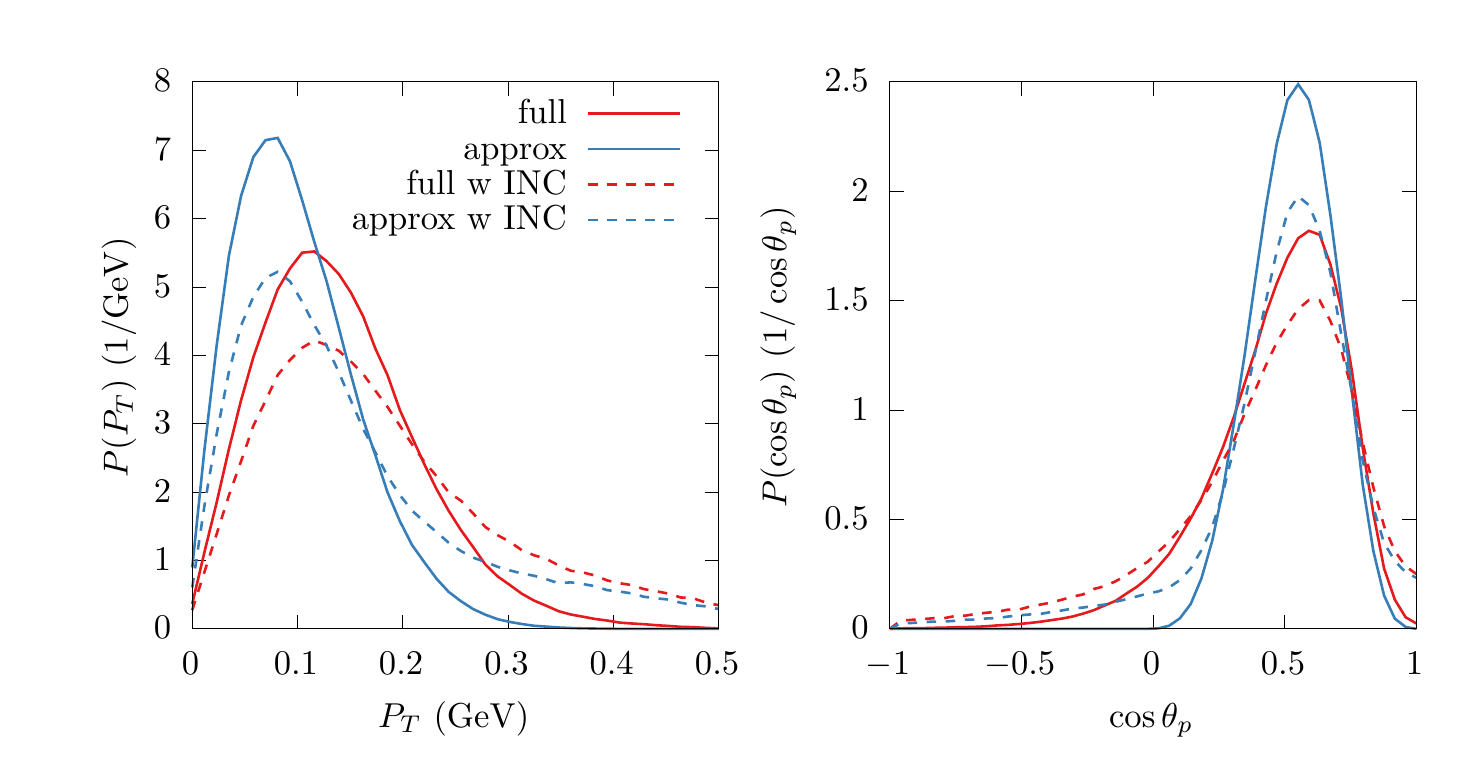}
\caption{Distributions of transverse momentum (left) and scattering angle with respect to the electron beam (right). Results for scattering of $1.159~\mathrm{GeV}$ electrons off carbon. The solid lines correspond to the same calculations as in Fig.~\ref{fig:TP_CTp_EDRMF}. The corresponding dashed lines use in addition the GENIE hN INC model. }
\label{fig:PT_CTp_EDRMF_hN}
\end{figure}

Given a vector $\vec{p}_m$ sampled from the momentum distribution, the outgoing nucleon energy is determined as
\begin{equation}
\label{eq:EnucGENIE}
E_N = \sqrt{p_m^2 + M_N^2} + \omega - E_{b}(q),
\end{equation}
if $\sqrt{p_m^2 + M_N^2} - M_N - E_b(q)$ is negative, otherwise a new $\vec{p}_m$ is generated.
Here $E_{b}(q)$ is a $q$-dependent binding energy inspired by the energy shift from the SuSAv2 model 
\begin{equation}
E_{b} = \max(5,-17.687 + 0.0564q) [\mathrm{MeV}] ~\left( q < 827~\mathrm{MeV}\right)
\end{equation}
and $E_b(q) = E_b(q=827~\mathrm{MeV})$ for $q > 827~\mathrm{MeV}$.
The resulting distributions of the nucleon kinetic energy are shown in the left panel of Fig.~\ref{fig:TP_CTp_EDRMF}. The procedure is found to lead to an overall shift of approximately $20~\mathrm{MeV}$ of the whole distribution. This could be amended by including an additional energy shift, which GENIE allows for. On the other hand the implementation of the on-shell dispersion relation in this expression comes from the FG approach and should be considered unrealistic~\cite{VOrden2019}.

The angular distributions (right panel of Fig.~\ref{fig:TP_CTp_EDRMF}) exhibit significant shape differences, partly due to the more restricted missing momentum distribution.
An approximation is made when determining the nucleon angle that also contributes to the narrower angular distribution. 
It is clear that the magnitude of the nucleon momentum generated with Eq.~\ref{eq:EnucGENIE} will not agree with momentum conservation, $\sqrt{E_N^2 - M_N^2} \neq \lvert \vec{p}_m + \vec{q} \rvert$.
To impose momentum conservation, the magnitude of the nucleon momentum is taken from Eq.~\ref{eq:EnucGENIE}, while the direction is taken from the momentum vectors, i.e. 
\begin{equation}
\label{eq:vecKnapprox}
\vec{k}_N = \sqrt{E_N^2 - M_N^2} \frac{\left(\vec{q} + \vec{p}_m \right)}{\lvert\vec{q} + \vec{p}_m \rvert}.
\end{equation}
The residual momentum is given to the remnant nucleus.

One might expect that these discrepancies are smeared out by rescattering in the INC, and by flux-folding in neutrino experiments. 
Indeed, in Ref.~\cite{Franco-Patino:2022tvv} it is seen that the SuSAv2 implementation is more similar to the EDRMF than found here, for flux-folded cross sections.
Some differences remain in the hadron distributions, but note that in Ref.~\cite{Franco-Patino:2022tvv} the SuSAv2 results include the INC, while the EDRMF ones do not.
We have computed the effect of rescattering by propagating the nucleon, both from the full EDRMF calculation and the GENIE procedure described above, through the nucleus using the $'hN'$ INC~\cite{Dytman:2021ohr}.
This results are shown in Fig.~\ref{fig:PT_CTp_EDRMF_hN}. The left panel shows the transverse momentum $P_T = \lvert \vec{k}_{e^\prime}^{T} + \vec{k}_{N}^{T} \rvert$, with ${}^{T}$ denoting the components orthogonal to the beam direction.
The angular distributions (right panel) are smeared out, but the effect of the approximation remains visible. For the $P_T$ distribution the differences are significant, and they affect the interpretation of experiments at fixed electron energy if this approximate treatment is used.

\section{Discussion}
In neutrino event generators, when only the inclusive cross section is known, an approximation is used to attach hadron variables to every event in order to describe semi-inclusive signals.
We have provided an overview of the algorithm used in GENIE, e.g. in the implementations of Refs.~\cite{Dolan19,Dolan:2021rdd}.
We have compared the nucleon observables that result from this algorithm using a local FG in quasielastic electron scattering, with calculations done in the relativistic distorted wave impulse approximation for kinematics relevant to the $e4\nu$ analysis~\cite{CLAS:2021neh}.
We find that the distribution of proton angles with respect to the beam and of the transverse momentum $P_T$ are significantly affected by this approximate treatment. We compute the effect of rescattering by using the $'hN'$ cascade model from GENIE, and find that differences remain large, in particular for $P_T$.

One can attribute part of these discrepancies to the use of the LFG momentum distribution. Indeed, while the LFG has been relatively successful in describing semi-inclusive flux-averaged neutrino cross sections~\cite{Bourguille:2020bvw}, the LFG spectral function should be considered unrealistic~\cite{VOrden2019} and hence fails under more restricted kinematic conditions. Additionally, due to the approximation made in selecting the nucleon angles Eq.~(\ref{eq:vecKnapprox}), the missing momentum distribution of this procedure will not correspond to the LFG momentum distribution used as input.
These issues can be amended by using instead a more realistic spectral function, e.g. Refs.~\cite{Gonzalez-Jimenez:2021ohu, BENHAR2005, Rocco:2018mwt}, and a modification to the selection of the angle which will be presented in future work.
Nonetheless, even in this case one does not obtain a fully consistent treatment suitable for every model, and in any case one loses the full kinematic and dynamical structure of the cross section~\cite{Moreno:2014}.

The RDWIA treatment used here, possibly with the inclusion of a spectral function beyond the mean field (see Ref.~\cite{Gonzalez-Jimenez:2021ohu}), does retain this structure and provides a way to study and benchmark such approximate treatments. Alternatively, the RDWIA events can be used directly in conjunction with a cascade model as done here and in Ref.~\cite{Nikolakopoulos:2022qkq}.
This approach will be used in future work to study the influence of nucleon distortion on the semi-inclusive cross section, and to establish the applicability and limitations of the cascade models used in neutrino event generators.

\vspace{6pt} 




\funding{ Fermilab is operated by the Fermi Research Alliance, LLC under contract No.~DE-AC02-07CH11359 with the United States Department of Energy. R.G.-J was supported by the government of Madrid and Complutense University under project PR65/19-22430}

\dataavailability{The event distributions used for this work, and a C++ program to compute alternative nucleon variables are available from A. Nikolakopoulos upon request.}

\conflictsofinterest{The authors declare no conflict of interest.}


%


\begin{adjustwidth}{-\extralength}{0cm}

\reftitle{References}


\bibliography{bibliography}

\end{adjustwidth}
\end{document}